\documentclass[aps,prl,twocolumn,superscriptaddress]{revtex4}
\usepackage{graphicx}
\usepackage{subfigure}
\usepackage{epstopdf}
\usepackage{amsmath}
\usepackage{amssymb}
\usepackage{amsfonts}
\usepackage{mathrsfs}
\usepackage{theorem}
\usepackage{bm}
\usepackage{url}
\usepackage[T1]{fontenc}
\usepackage{csquotes}
\MakeOuterQuote{"}

\usepackage{algorithm}
\usepackage{algorithmicx}
\usepackage{algpseudocode}

\usepackage{dcolumn}
\usepackage{color}

\definecolor{ngreen}{rgb}{0.2,0.6,0.2}

\definecolor{ngold}{rgb}{0.7,0.6,0.2}

\def\<{\langle}  
\def\>{\rangle}  

\def\vec#1{\mathbf{#1}} 

\newcommand{\tr}{\operatorname{tr}}
\newcommand{\Tr}{\operatorname{Tr}}

\newcommand{\rmi}{\mathrm{i}}
\newcommand{\rme}{\mathrm{e}}

\newcommand{\rmW}{\mathrm{W}}

\newcommand{\be}{\begin{equation}}
\newcommand{\ee}{\end{equation}}
\newcommand{\ba}{\begin{align}}
\newcommand{\ea}{\end{align}}

\def\<{\langle}  
\def\>{\rangle}  
\newcommand{\ket}[1]{| #1\>}
\newcommand{\bra}[1]{\< #1|}

\newcommand{\inner}[2]{\<#1|#2\>}
\def\outer#1#2{|#1\>\<#2|}       

\def\eqref#1{\textup{(\ref{#1})}}  
\newcommand{\eref}[1]{Eq.~\textup{(\ref{#1})}}

\newcommand{\esref}[1]{Eqs.~\textup{(\ref{#1})}}

\newcommand{\fref}[1]{Fig.~\ref{#1}}

\newcommand{\fsref}[1]{Figs.~\ref{#1}}

\newcommand{\tref}[1]{Table~\ref{#1}}

\newcommand{\cref}[1]{Conjecture~\ref{#1}}
\newcommand{\Cref}[1]{Conjecture~\ref{#1}}

\newcommand{\rcite}[1]{Ref.~\cite{#1}}

\begin{document}

\title{Experimental optimal orienteering via parallel and antiparallel spins}
\author{Jun-Feng Tang}
\thanks{These authors contributed equally to this work.}
\affiliation{Key Laboratory of Quantum Information,University of Science and Technology of China, CAS, Hefei 230026, P. R. China}
\affiliation{Synergetic Innovation Center of Quantum Information and Quantum Physics, University of Science and Technology of China, Hefei 230026, P. R. China}
\author{Zhibo Hou}
\thanks{These authors contributed equally to this work.}
\affiliation{Key Laboratory of Quantum Information,University of Science and Technology of China, CAS, Hefei 230026, P. R. China}
\affiliation{Synergetic Innovation Center of Quantum Information and Quantum Physics, University of Science and Technology of China, Hefei 230026, P. R. China}
\author{Jiangwei Shang}
\email{jiangwei.shang@bit.edu.cn}
\affiliation{Key Laboratory of Advanced Optoelectronic Quantum Architecture and Measurement of Ministry of Education, School of Physics, Beijing Institute of Technology, Beijing 100081, China}
\author{Huangjun~Zhu}
\email{zhuhuangjun@fudan.edu.cn}

\affiliation{Department of Physics and Center for Field Theory and Particle Physics, Fudan University, Shanghai 200433, China}

\affiliation{State Key Laboratory of Surface Physics, Fudan University, Shanghai 200433, China}

\affiliation{Institute for Nanoelectronic Devices and Quantum Computing, Fudan University, Shanghai 200433, China}

\affiliation{Collaborative Innovation Center of Advanced Microstructures, Nanjing 210093, China}
\author{Guo-Yong Xiang}
\email{gyxiang@ustc.edu.cn}
\affiliation{Key Laboratory of Quantum Information,University of Science and Technology of China, CAS, Hefei 230026, P. R. China}
\affiliation{Synergetic Innovation Center of Quantum Information and Quantum Physics, University of Science and Technology of China, Hefei 230026, P. R. China}
\author{Chuan-Feng Li}
\affiliation{Key Laboratory of Quantum Information,University of Science and Technology of China, CAS, Hefei 230026, P. R. China}
\affiliation{Synergetic Innovation Center of Quantum Information and Quantum Physics, University of Science and Technology of China, Hefei 230026, P. R. China}
\author{Guang-Can Guo}
\affiliation{Key Laboratory of Quantum Information,University of Science and Technology of China, CAS, Hefei 230026, P. R. China}
\affiliation{Synergetic Innovation Center of Quantum Information and Quantum Physics, University of Science and Technology of China, Hefei 230026, P. R. China}

\begin{abstract}
Antiparallel spins are superior in orienteering to parallel spins. This intriguing phenomenon is tied to entanglement associated with quantum measurements rather than quantum states. Using photonic systems, we experimentally realize the optimal orienteering protocols based on parallel spins and antiparallel spins, respectively. The optimal entangling measurements for decoding the direction information from parallel spins and antiparallel spins are realized using  photonic quantum walks, which is a useful idea that is  of wide interest in quantum information processing and foundational studies. Our experiments clearly demonstrate the advantage of antiparallel spins over parallel spins in orienteering. In addition, 
entangling measurements can extract more information than local measurements even if no entanglement is present in the quantum states. 
\end{abstract}


\date{\today}
\maketitle

\textit{Introduction.---}Quantum information processing promises to realize many tasks, such as computation, communication, and metrology \cite{NielC00book,Gision02quantum,GiovLM11}, much more efficiently than the classical counterpart. The power of quantum information processing is closely tied to quantum entanglement  \cite{Horo09quantum,Guehne.Toth2009}, the characteristic feature  of quantum mechanics. Entanglement can manifest in both quantum states and quantum measurements \cite{MassP95,GisiP99,BagaBGM06S,VidrDGJ14,RoccGMS17,ZhuH17U}, and the former    has been extensively studied in the past thirty years. By contrast, entanglement in quantum measurements is still not well understood \cite{gisin2019entanglement}, although it is connected to a number of intriguing phenomena, such as "nonlocality without entanglement" \cite{BennDFM99}.

\begin{figure*}[t]
	\center{\includegraphics[scale=0.85]{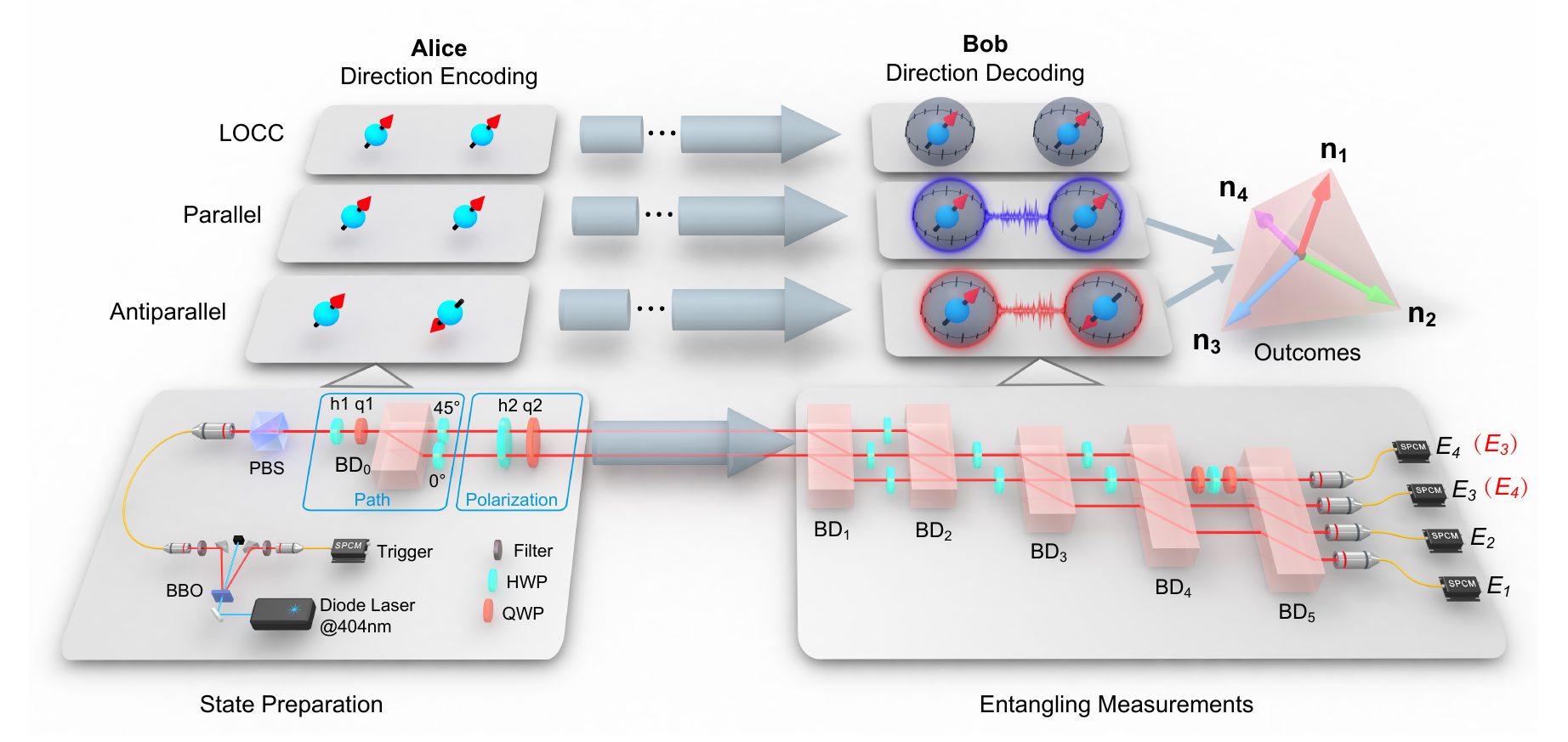}}
	\caption{\label{fig: exp setup}   Schematic diagram and experimental setup for optimal orienteering with parallel and antiparallel spins. 
	Direction encoding on Alice's side is implemented in the module of ``State Preparation'', which prepares the two (parallel  or antiparallel) spins in path and polarization  degrees of freedom, respectively. After receiving the two spins, Bob decodes the direction information using the optimal entangling measurement realized via photonic quantum walks. 
  Here a polarizing beam splitter (PBS) initializes the polarization state in H-component, and beam displacers  (BDs)  realize the conditional translation operator $T$.  Half wave plates (HWPs) and quarter wave plates (QWPs)  realize site-dependent  coin operators $C(x,t)$.	Four single-photon-counting modules (SPCMs) $E_1$ to $E_4$  correspond to the four outcomes of the entangling measurement. Note that  the positions of  $E_3$ and $E_4$ are switched in the case of antiparallel decoding, as marked in red. 
	}
\end{figure*}

A classical task for which entangling measurements play a central role is orienteering (direction encoding and decoding) using parallel and antiparallel spins \cite{MassP95,GisiP99,Mass00collective,Bart07reference}, first recognized by Gisin and Popescu twenty years ago \cite{GisiP99} (see \fref{fig: exp setup}). Suppose Alice wants to  communicate a  random space direction $\vec{n}$  to Bob and she can send only two spin-1/2 particles. A natural way to encode the direction is to polarize the two spins  along the same direction $\vec{n}$, as characterized by the ket $|\vec{n},\vec{n}\>$. 
After receiving the two spins, Bob can perform some measurement and guess the direction based on the measurement outcome. The performance of Bob is characterized by the average fidelity of his guess and the original spin state.  Alternatively, Alice may send two spins polarized along opposite directions, that is,  $|\vec{n},-\vec{n}\>$.

In either way, there is no entanglement between the two spins and, intuitively, one will not expect any advantage of one strategy over the other. This conclusion indeed holds if Bob's measurement on the two spins  requires only local operations and classical communication (LOCC), in which case the maximum fidelity Bob can achieve is  $(3+\sqrt{2})/6\approx0.7357$ for both encoding methods \cite{Mass00collective}. However, the situation is different if Bob can perform entangling measurements. Now, the maximum fidelity  is $3/4=0.75$ for the parallel encoding and  $(3+\sqrt{3})/6\approx 0.7887$ for the antiparallel encoding~\cite{GisiP99}. This intriguing phenomenon manifests the importance of entanglement in quantum measurements instead of quantum states. 
Although this canonical example is well known by now, no convincing experimental demonstration is known to us in the literature. Incidentally, in  the experiment reported in Ref.~\cite{Jeff06optical}, the entanglement was mapped to the state preparation process instead, which contradicts the spirit of the original proposal and is thus hardly convincing for demonstrating the power of entangling measurements.

%


 
%
%
%

Using photonic systems here we realize optimal orienteering with  parallel spins and antiparallel spins.
The optimal protocol based on LOCC is also realized as a benchmark. To achieve this goal, we encode the two spins into polarization and path degrees of freedom of a photon, respectively. Then the optimal measurements are realized using photonic quantum walks. Measurement tomography shows that these measurements are realized with high qualities. The optimal fidelities we achieved agree very well with the theoretical predictions. These results demonstrate convincingly  that
antiparallel encoding is indeed better than parallel encoding for communicating the direction. Also, 
entangling measurements are more efficient than separable measurements for extracting the direction information.
Our work is expected to stimulate more researches on quantum entanglement in measurements, which deserves much further studies.

\textit{Optimal measurements for two spins.}---%
In the original paper \cite{GisiP99}, Gisin and Popescu considered the communication of a completely random direction. Simple analysis shows that all the conclusions remain the same if the direction $\vec{n}$ is chosen \emph{a priori} on the vertices of a regular octahedron with equal probability of $1/6$. 
This  simpler setting is more appealing to demonstrate 
the distinction between  parallel encoding  and antiparallel encoding. 

Suppose Alice chooses the direction $\vec{n}=(x,y,z)$ with $x^2+y^2+z^2=1$ at random (uniformly either from the unit sphere or the vertices of the regular octahedron) and encodes it into two parallel spins $|\vec{n},\vec{n}\>$, where $|\vec{n}\>$  is a qubit ket with Bloch vector $\vec{n}$
and density matrix $\rho=|\vec{n}\>\<\vec{n}|=(1+\vec{n}\cdot \bm{\sigma})/2$.  Here $\bm{\sigma}$ is the vector composed of the three Pauli matrices $\sigma_x, \sigma_y,\sigma_z$. If Bob can only access LOCC, then the optimal protocol after receiving the two spins can be realized as follows \cite{Mass00collective}. Bob first measures one spin along some direction $\vec{a}$ and then measures the other spin along an orthogonal direction $\vec{b}$. Denote the outcomes of the two measurements by $\pm \vec{a}$ and $\pm\vec{b}$, respectively, then the guess direction is the bisectrix of the two vectors associated with the two outcomes. 
For a given $\vec{n}$, the mean fidelity achieved by this protocol is
\begin{equation}\label{eq:LOCCmF}
\frac{1}{4}\bigl[2+\sqrt{2}(\vec{n}\cdot \vec{a})^2+\sqrt{2}(\vec{n}\cdot \vec{b})^2\bigr]. 
\end{equation}
The average fidelity over uniform distribution on the sphere or on the vertices of the octahedron is 
about 0.7357,  which achieves the maximum under LOCC \cite{GisiP99,Mass00collective}. To  be concrete, Bob can measure the pair $\sigma_x, \sigma_y$ on the two spins, respectively; pairs $\sigma_z, \sigma_x$ and $\sigma_z, \sigma_y$ are equally good (see  \tref{tab:dictionary} in the supplement).

If Bob can access entangling measurements, then the optimal protocol is realized by the projective measurement onto the  basis composed of the four states
\cite{chang2014optimal}
\begin{equation}\label{eq:parallel spins}
\ket{\Psi_j^{\parallel}} =\frac{\sqrt{3}}{2}\ket{\vec{n}_j,\vec{n}_j}+\frac{1}{2}\ket{\Psi_-},\quad j=1,2,3,4,
\end{equation}
where $|\Psi_-\>=\frac{1}{\sqrt{2}}(|01\>-|10\>)$ is the singlet, which is maximally entangled, and $\ket{\vec{n}_j}$ for $j=1,2,3,4$ are qubit states that form a
symmetric informationally complete positive operator-valued measure (SIC POVM), that is, $|\<\vec{n}_j|\vec{n}_k\>|^2=(2\delta_{jk}+1)/3$ \cite{Zaun11,ReneBSC04}. Geometrically, the Bloch vectors $\vec{n}_j$ form a regular tetrahedron inside the Bloch sphere. To make sure that  the four states in \eref{eq:parallel spins} are orthogonal, we can choose
\begin{equation}\label{eq:QubitSIC}
\begin{aligned}
\ket{\vec{n}_1}=&\ket{0},\qquad \ket{\vec{n}_2}=\frac{\rmi}{\sqrt{3}}(\ket{0}+\sqrt{2}\ket{1}), \\ \ket{\vec{n}_3}=&\frac{\rmi}{\sqrt{3}}(\ket{0}+\rme^{\frac{2\pi}{3}\rmi}\sqrt{2}\ket{1}),\\ \ket{\vec{n}_4}=&\frac{\rmi}{\sqrt{3}}(-\ket{0}+\rme^{\frac{\pi}{3}\rmi}\sqrt{2}\ket{1}).
\end{aligned}
\end{equation}
The guess direction is $\vec{n}_j$ if outcome  $j$ in \eref{eq:parallel spins} appears upon the measurement.
For a given $\vec{n}$, the mean fidelity achieved by this protocol is
\begin{equation}\label{eq:PmF}
\frac{1}{24}(18+\sqrt{2} x^3-3\sqrt{2} xy^2-3x^2z-3y^2z+2z^3).
\end{equation}
The average of this fidelity over any  distribution of $\vec{n}$ that is symmetric under inversion is 0.75. In particular, the average over uniform distribution on the sphere or on the vertices of the octahedron is 
$0.75$,  which achieves the maximum for parallel encoding \cite{GisiP99,Mass00collective}.

Next, suppose Alice encodes the direction $\vec{n}$ into antiparallel spins $|\vec{n},-\vec{n}\>$. Now, the optimal protocol can be realized by performing the projective measurement on the basis
\begin{equation}\label{eq:antiparallel spins}
\ket{\Psi_j^{\perp}} =\frac{\sqrt{3}+1}{2\sqrt{2}}\ket{\vec{n}_j,-\vec{n}_j}+\frac{\sqrt{3}-1}{2 \sqrt{2}}\ket{-\vec{n}_j,\vec{n}_j},
\end{equation}
where $\ket{\vec{n}_j}$ are given in  \eref{eq:QubitSIC} and $\ket{-\vec{n}_j}$ are chosen as follows,
\begin{equation}\label{eq:QubitSICm}
\begin{aligned}
\ket{-\vec{n}_1}=&\ket{1},\qquad \ket{-\vec{n}_2}=\frac{\rmi}{\sqrt{3}}(\sqrt{2}\ket{0}-\ket{1}), \\ \ket{-\vec{n}_3}=&\frac{\rmi}{\sqrt{3}}(\rme^{-\frac{2\pi}{3}\rmi}\sqrt{2}\ket{0}-\ket{1}),\\ \ket{-\vec{n}_4}=&\frac{\rmi}{\sqrt{3}}(\rme^{-\frac{\pi}{3}\rmi}\sqrt{2}\ket{0}+\ket{1}).
\end{aligned}
\end{equation}
The guess direction is $\vec{n}_j$ if outcome  $j$  in \eref{eq:antiparallel spins} appears. For a given $\vec{n}$, the mean fidelity achieved  is
\begin{equation}\label{eq:APmF}
\frac{1}{12}(6+2\sqrt{3}+\sqrt{2} x^3-3\sqrt{2} xy^2-3x^2z-3y^2z+2z^3).
\end{equation}
The average  over any inversion-symmetric  distribution is about 0.7887 \cite{GisiP99,Mass00collective}, which is larger than the counterpart for parallel encoding, though  the fluctuation is larger.  Incidentally, the measurement defined in \eref{eq:antiparallel spins} was called the Elegant Joint Measurement by Gisin and plays an important role in the study of  N-locality  \cite{Gisi17,gisin2019entanglement}.


\bigskip
\textit{Realization of the optimal measurements via quantum walks.}---%
Quantum walks are a powerful tool in quantum information processing, including quantum computation and quantum simulation. Recently,  quantum walks also found important applications in implementing generalized  measurements 
 \cite{KurzW13, BianLQZ15, ZhaoYKX15, LiZZ19}. Consider  a  quantum walk on a one-dimensional chain, and the system  is characterized by two degrees of freedom $|x,c\>$, where $x$ denotes the walker position and can take any integer value, while $c=0,1$ denotes the coin state. The evolution in  each step is 
determined by a unitary transformation of the form $U(t)=TC(t)$, where
\begin{align}\label{eq:translation operator}
  T = \sum_x \ket{x+1,0}\bra{x,0} + \ket{x-1,1}\bra{x,1}
  \end{align}
is the conditional translation operator,
and $C(t)=\sum_x |x\>\<x|\otimes C(x,t)$ 
is determined by site-dependent coin operators $C(x,t)$. Any discrete
 POVM on a qubit can be realized by choosing suitable  coin operators $C(x,t)$ and then measuring the walker position after sufficiently many steps \cite{KurzW13}. In addition, quantum walks can be used to realize POVMs on higher-dimensional systems  \cite{LiZZ19}, including collective  measurements on a two-qubit system \cite{hou2018deterministic}.


 Here we use quantum walks to realize 
 optimal entangling measurements for decoding the spin direction from  parallel encoding  and antiparallel encoding, as specified in \esref{eq:parallel spins} and \eqref{eq:antiparallel spins}. 
 To realize these two-qubit projective measurements using quantum walks, we take the coin qubit and the walker in positions 1 and $-1$  as the two-qubit system of interest and use  other positions of the walker as an ancilla.
In this way, the two-qubit projective measurements  in \esref{eq:parallel spins} and \eqref{eq:antiparallel spins} can be realized with five-step photonic quantum walks as shown in the  module of ``Entangling Measurements'' in \fref{fig: exp setup}.  At each step, the state of the coin qubit is transformed by the coin operator $C(x,t)$ depending on the walker position.  Upon the action of the translation operator, then the position of the walker is updated based on
 the coin state. After certain steps, measurement of the walker position effectively realizes a POVM  (including projective measurements) on the two-qubit system composed of the walker and coin. In particular, we can realize the optimal entangling measurements in \eref{eq:parallel spins} and  \eref{eq:antiparallel spins} with five-step quantum walks by designing the coin operators $C(x,t)$ wisely (see the supplement).
The four detectors $E_1$ to $E_4$ marked in the figure correspond to the four projectors onto the four basis states $\ket{\Psi_1^{\parallel}}$ to $\ket{\Psi_4^{\parallel}}$  tailored for parallel encoding and $\ket{\Psi_1^{\perp}}$ to  $\ket{\Psi_4^{\perp}}$ tailored for antiparallel encoding. This setup can also be used to realize local projective measurements  $\sigma_x\sigma_y, \sigma_z\sigma_x$, and  $\sigma_z\sigma_y$, which are optimal under LOCC.




\bigskip
\textit{Experimental setup.}---%
The experimental setup for optical orienteering via parallel and antiparallel encodings as well as decoding with entangling measurements is illustrated in \fref{fig: exp setup}. The setup is composed of  two modules designed for state preparation of  parallel (antiparallel) spins  and entangling measurements, respectively.



 \begin{figure*}
	\center{\includegraphics[scale=0.45]{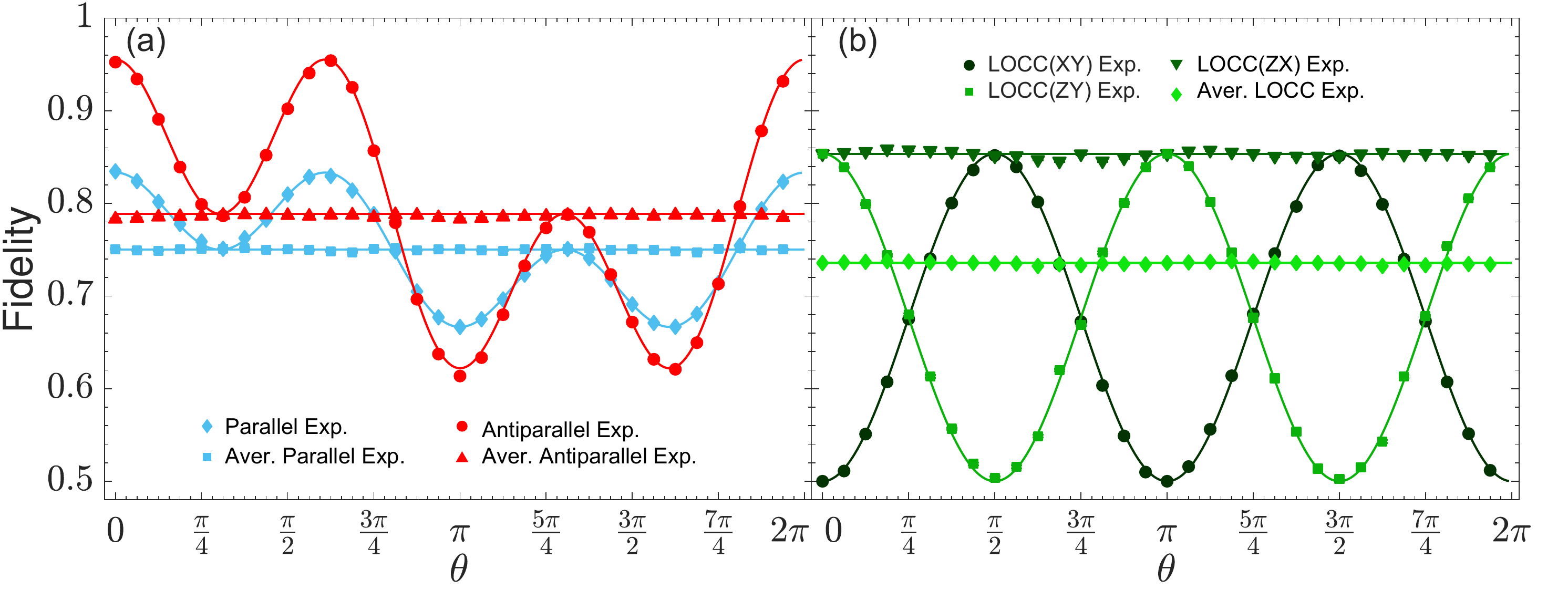}}
	\caption{\label{fig: fidelity} Fidelities  of transferring a class of directions $(\sin\theta,0,\cos\theta)$
	based on parallel and antiparallel spins.  (a) Performances of optimal entangling measurements on parallel and antiparallel spins; (b) Performances of local projective  measurements on parallel spins. Each data point is the average over  50000 runs. To manifest the direction-independent behavior, the  fidelities averaged over directions $\theta$ and $\theta+\pi$ are also shown in plot (a); by contrast,
the fidelities  averaged over three local projective  measurements are shown in plot (b).
	 The error bar denotes the standard deviation of 100 numerical simulations from Poisson statistics. }
\end{figure*}


In the  module of ``State Preparation'',  Alice encodes the desired direction $\vec{n}$ into the Bloch vectors of qubit 1  and qubit 2 in the path and polarization degrees of freedom, i.e., the walker qubit encoded in positions 1 and $-1$ and the coin qubit with H and V polarizations. A  2-mm-long BBO crystal, cut for type-\uppercase\expandafter{\romannumeral1} phase-matched spontaneous parametric down-conversion (SPDC) process, is pumped by a 40-mW V-polarized beam at 404~nm.  After the SPDC process,  a pair of photons with wavelength $\lambda=808$~nm are created in the state of $\ket{HH}$ \cite{Kwia99ultrabright}. The two photons pass through two interference filters with a bandwidth of 3 nm. The  two-photon coincidence counts are about 7000 per second.  One photon is detected by a single-photon-counting module acting as a trigger. The other photon acts as a heralding single-photon source and is prepared in $\ket{H}$ by a polarizing beam splitter (PBS).  The desired direction $\ket{\vec{n}}$ is encoded  in the Bloch vector of the photon by a HWP  and a QWP  with deviation angles $h_1, q_1$.  To transform the polarization state into the path state,  BD$_0$ is used to displace the H-component and V-component into two paths; then  a HWP with deviation angle  $45^\circ$ is placed in the V-component path to prepare the photon in the state $\ket{\vec{n},H}$.

 Next, Alice encodes the ket $\ket{\vec{n}}$ or $\ket{-\vec{n}}$ into the polarization degree of freedom  (coin qubit) using a HWP and a QWP. In this way,  Alice can prepare the desired parallel spins $\ket{\vec{n},\vec{n}}$ or antiparallel spins $\ket{\vec{n},-\vec{n}}$, the first qubit of which is encoded in the path degree of freedom, while the second one  in the polarization degree of freedom.

Then, the two-spin state  is sent into the  module of ``Entangling Measurements'' on Bob's side, which performs the entangling measurements in \eref{eq:parallel spins} or \eqref{eq:antiparallel spins} based on quantum walks; see \fref{fig:para scheme} in the supplement for more details. To realize  the conditional translation operator $T$, we use  interferometrically stable beam displacers  (BDs) \cite{Obr03demonstration,Rah13direct,Rab17entanglement,Sha15strong} to separate horizontal polarization (H)  4 mm away from vertical polarization (V). Each coin operator is realized by no more than three  half wave plates (HWPs) or quarter wave plates (QWPs).  The rotation angles are specified in  the table within  \fref{fig:para scheme} in the supplement.
According to  the measurement scheme and its outcome, Bob guesses a direction $\vec{n}_g$ by virtue of  the dictionary in \tref{tab:dictionary}. 
To accurately characterize the optimal entangling measurements as well as local projective measurements
that were actually realized, we performed quantum measurement tomography \cite{Fiur01maximum} 
and demonstrated that these measurements were experimentally realized with very high fidelities (see the supplement).


 

\begin{table*}[t]
	\caption{\label{tab:MUB fideltiy} Fidelities of transferring six directions corresponding to the vertices of the regular octahedron.  Two  entangling measurements for parallel and antiparallel spins and three local projective  measurements  are compared.
	Each data point is the average over  50000 runs.	The number in the parentheses indicates the standard deviation of 100 numerical simulations from Poisson statistics.
	}
	\begin{tabular*}{1\textwidth}{@{\extracolsep{\fill}} c| c c c c c c c}
		\hline\hline
		Measurement schemes & $(1,0,0)$& $(-1,0,0)$&$(0,1,0)$ & $(0,-1,0)$&$(0,0,1)$ & $(0,0,-1)$& average\\
		\hline
		parallel&0.8018(11)&0.6919(4)&0.7494(8)&0.7492(9)&0.8417(13)&0.6665(1)&0.7501(4)\\
		antiparallel&0.9023(7)&0.6713(7)&0.7847(8)&0.7905(9)&0.9541(8)&0.6142(8)&0.7862(3)\\
		\hline
		$\sigma_x\sigma_y$&0.8472(3)&0.8534(1)&0.8530(0)&0.8526(1)&0.5000(0)&0.5000(0)&0.7344(1)\\
		$\sigma_z\sigma_x$&0.8512(2)&0.8514(2)&0.5000(0)&0.5000(0)&0.8535(1)&0.8534(1)&0.7349(1)\\
		$\sigma_z\sigma_y$&0.5000(0)&0.5000(0)&0.8518(1)&0.8512(2)&0.8535(1)&0.8535(1)&0.7350(1)\\
		\hline\hline
	\end{tabular*}
\end{table*}



%
%
%

%


\bigskip
\textit{Optimal orienteering via parallel and antiparallel spins.}---%
By virtue of the optimal entangling measurements realized using quantum walks, we can now demonstrate the distinction between parallel spins and antiparallel spins for orienteering.

%
%



First, we verify the fidelity formulas presented in \esref{eq:PmF} and \eqref{eq:APmF}. In the experiment, Alice draws a direction vector on the $xz$ plane, which has the form $\vec{n}=(\sin\theta,0,\cos\theta)$ with  $0\leq \theta\leq 2\pi$ (this information is hidden from Bob) and applies parallel encoding $|\vec{n},\vec{n}\>$ or antiparallel encoding $|\vec{n},-\vec{n}\>$, where $|\vec{n}\>=\cos\frac{\theta}{2}\ket{0}+\sin\frac{\theta}{2}\ket{1}$ and $|-\vec{n}\>=-\sin\frac{\theta}{2}\ket{0}+\cos\frac{\theta}{2}\ket{1}$. 
After receiving the  two qubits which encode the direction information, Bob 
performs the optimal entangling measurement (depending on the encoding method of Alice) and guess the direction  $\vec{n}_g$  using the dictionary in \tref{tab:dictionary} in the supplement. The fidelity of his guess is defined as $F=(1+\vec{n}\cdot\vec{n}_g)/2$, and the average fidelity over 50000 runs for each strategy is shown in \fref{fig: fidelity}, which
agrees very well with the theoretical predication. Notably, the average fidelity averaged over antipodal points $\theta$ and $\theta+\pi$ is almost independent of  $\theta$ for both encoding methods as predicted; in addition, the average fidelity for antiparallel encoding is clearly larger than that for parallel encoding. As a benchmark, in the case of parallel encoding, we also considered the scenario in which Bob performs local projective measurements on the two qubits separately.

Next, Alice draws one of the six directions $\pm x, \pm y,\pm z$ at random and apply parallel or antiparallel encoding.  
After receiving the  two qubits which encode the direction information, Bob can perform one of the five measurement schemes, three of which are optimal local projective measurements, while the other two are optimal entangling measurements tailored for parallel encoding  and antiparallel encoding, respectively. Based on 
the measurement outcome, Bob makes his guess $\vec{n}_g$,  and the average fidelity over 50000 runs for each strategy is shown in \tref{tab:MUB fideltiy}. The experimental results closely match the theoretical maximums achievable by LOCC  (0.7357), optimal measurements for parallel encoding  (0.75), and optimal measurements for antiparallel encoding (0.7887), respectively. In this way, our experiment clearly demonstrates that antiparallel encoding can achieve better orienteering than parallel encoding. 
Meanwhile, entangling measurements are more powerful in extracting the direction information than local measurements.

\bigskip
\textit{Summary.}---%
Using photonic quantum walks, we experimentally realized the optimal entangling measurements for decoding the direction from parallel spins and antiparallel spins, respectively. Our experiments clearly demonstrate that  antiparallel spins are superior to parallel spins in orienteering. In addition, 
entangling measurements can extract more direction information than local measurements. Although it is difficult to realize practical orienteering using the current proposal, our work represents an important step in exploring the power of entangling measurements in quantum information processing as well as foundational studies, and is thus expected to stimulate more researches on entangling measurements. 
In particular, the optimal measurement on antiparallel spins realized in our experiments is also of key interest in the study of N-locality \cite{Gisi17,gisin2019entanglement}.


%


%
\acknowledgments
The work at USTC is supported by the National Natural Science Foundation of China under Grants (Nos. 11574291, 11774334, 61327901 and 11774335),  the National Key Research and Development Program of China (No.2017YFA0304100),  Key Research Program of Frontier Sciences, CAS (No.QYZDY-SSW-SLH003), the Fundamental Research Funds for the Central Universities (No.WK2470000026).  JS acknowledges support by the Beijing Institute of Technology Research Fund Program for Young Scholars and the National Natural Science Foundation of China (Grant No. 11805010). HZ is  supported by  the National Natural Science Foundation of China (Grant No. 11875110).



\clearpage
\newpage
\addtolength{\textwidth}{-1in}
\addtolength{\oddsidemargin}{0.5in}
\addtolength{\evensidemargin}{0.5in}

\setcounter{equation}{0}
\setcounter{figure}{0}
\setcounter{table}{0}

\makeatletter
\renewcommand{\theequation}{S\arabic{equation}}
\renewcommand{\thefigure}{S\arabic{figure}}
\renewcommand{\thetable}{S\arabic{table}}


\onecolumngrid
\begin{center}
	\textbf{\large Experimental optimal orienteering via parallel and antiparallel spins: Supplement}
\end{center}

In this supplement, we provide the  dictionary  of guessed directions $\vec{n}_g$ from measurement outcomes
in orienteering with parallel and antiparallel spins. We then give more details on the realization of the optimal entangling measurements and local projective measurements based on quantum walks. Finally, we present more details on the measurement tomography of these optimal measurements.

\section{The dictionary of guessed directions $\vec{n}_g$ from measurement outcomes.}

\begin{table*}[h]
	\caption{\label{tab:dictionary} The dictionary of guessed directions $\vec{n}_g$ from measurement outcomes. Here the left column lists the two optimal entangling measurements and three  optimal local projective measurements. $E_1$ to $E_4$ represent the four outcomes of each measurement scheme. Each vector in the table represents the guessed direction associated with the measurement indicated in the first column and the outcome indicated in the first row. 
	}
	\begin{tabular*}{1\textwidth}{@{\extracolsep{\fill}} c| c c c c}
		\hline\hline
		Measurement schemes & $E_1$& $E_2$&$E_3$ & $E_4$\\
		\hline
		parallel&$(0,0,1)$&$(2\sqrt{2},0,-1)/3$&$(-\sqrt{2},\sqrt{6},-1)/3$&$(-\sqrt{2},-\sqrt{6},-1)/3$\\
		antiparallel&$(0,0,1)$&$(2\sqrt{2},0,-1)/3$&$(-\sqrt{2},\sqrt{6},-1)/3$&$(-\sqrt{2},-\sqrt{6},-1)/3$\\
		\hline
		$\sigma_x\sigma_y$&$(1,1,0)/\sqrt{2}$&$(-1,-1,0)/\sqrt{2}$&$(1,-1,0)/\sqrt{2}$&$(-1,1,0)/\sqrt{2}$\\
		$\sigma_z\sigma_x$&$(1,0,1)/\sqrt{2}$&$(-1,0,1)/\sqrt{2}$&$(1,0,-1)/\sqrt{2}$&$(-1,0,-1)/\sqrt{2}$\\
		$\sigma_z\sigma_y$&$(0,-1,1)/\sqrt{2}$&$(0,1,1)/\sqrt{2}$&$(0,-1,-1)/\sqrt{2}$&$(0,1,-1)/\sqrt{2}$\\
		\hline\hline
	\end{tabular*}
\end{table*}

\section{Optimal entangling and local projective measurements via quantum walks}

In this section we propose concrete schemes based on quantum walks to realize optimal entangling measurements on parallel spins and antiparallel spins as well as optimal local projective measurements $\sigma_x\sigma_y, \sigma_z\sigma_x$, and $\sigma_z\sigma_y$.

Recall that  a  quantum walk on a one-dimensional chain is characterized by two degrees of freedom $|x,c\>$, where $x$ denotes the walker position, while $c=0,1$ denotes the coin state. The evolution in  each step is 
determined by a unitary transformation of the form $U(t)=TC(t)$, where
\begin{align}
T = \sum_x \ket{x+1,0}\bra{x,0} + \ket{x-1,1}\bra{x,1}
\end{align}
is the conditional translation operator,
and $C(t)=\sum_x |x\>\<x|\otimes C(x,t)$ 
is determined by site-dependent coin operators $C(x,t)$. After $k$ steps, the unitary operator generated by the coin operators and translation operator reads  $U=TC(k)\cdots TC(2)TC(1)$. Measurement of the walk position then effectively realizes a POVM on the coin qubit \cite{KurzW13,LiZZ19}. To see this, suppose initially  the coin state is $|\varphi\>$, and the walker is at position 0. Then the probability of finding the walker at position $j$ after $k$ steps reads
\begin{equation}
p_j =\tr [U(|0\>\<0|\otimes |\varphi\>\<\varphi|)U^\dag (|j\>\<j|\otimes\openone )]=\tr(\Pi_j |\varphi\>\<\varphi|). 
\end{equation}
Here  $\Pi_j$ is the POVM element corresponding to the  position $x=j$ and has the form
\begin{equation}\label{eq:Omega}
\Pi_j=\Tr_{\rmW}\{(|0\>\<0|\otimes\openone)U^{\dagger} \left(|j\rangle\langle j| \otimes \openone \right) U\},
\end{equation}
where "$\Tr_{\rmW}$" denotes the partial trace on the walker system.

 What is not so obvious is that any discrete
POVM on a qubit can be realized by choosing suitable  coin operators $C(x,t)$ and then measuring the walker position after sufficiently many steps \cite{KurzW13}. Moreover, the above scheme based on quantum walks can be generalized to realize any discrete POVM on a qudit  \cite{LiZZ19}. In addition, another variant  can be applied to implementing collective  measurements on a two-qubit system  composed of the coin and part of the walker degree of freedom \cite{hou2018deterministic}. The original measurement strategy presented in \rcite{hou2018deterministic} was based on ad hoc construction. Fortunately, a general algorithm for implementing  collective measurements was devised  recently by our collaborator Zihao Li (the first author of \rcite{LiZZ19}).


Here we use quantum walks to realize
two optimal entangling measurements and three local projective measurements presented in the main text for decoding the spin direction from  parallel encoding  and antiparallel encoding. 
To realize these two-qubit projective measurements using quantum walks, we take the coin qubit and the walker in positions 1 and $-1$  as the two-qubit system of interest and use  other positions of the walker as an ancilla.
 At each step, the state of the coin qubit is transformed by the coin operator $C(x,t)$ depending on the walker position.  Upon the action of the translation operator, then the position of the walker is updated based on
the coin state. After certain steps, measurement of the walker position effectively realizes a POVM  (including projective measurements) on the two-qubit system composed of the walker and coin. The specific POVM elements can be determined by analyzing the evolution of the walker-coin system under the actions of the coin operators and translation operator. In this way,  we can realize the optimal entangling measurements and local projective measurements mentioned above using five-step quantum walks by designing the coin operators $C(x,t)$ wisely as explained as follows.

\subsection{Optimal entangling measurement for parallel spins}

\begin{figure}[b]
	\center{\includegraphics[scale=0.6]{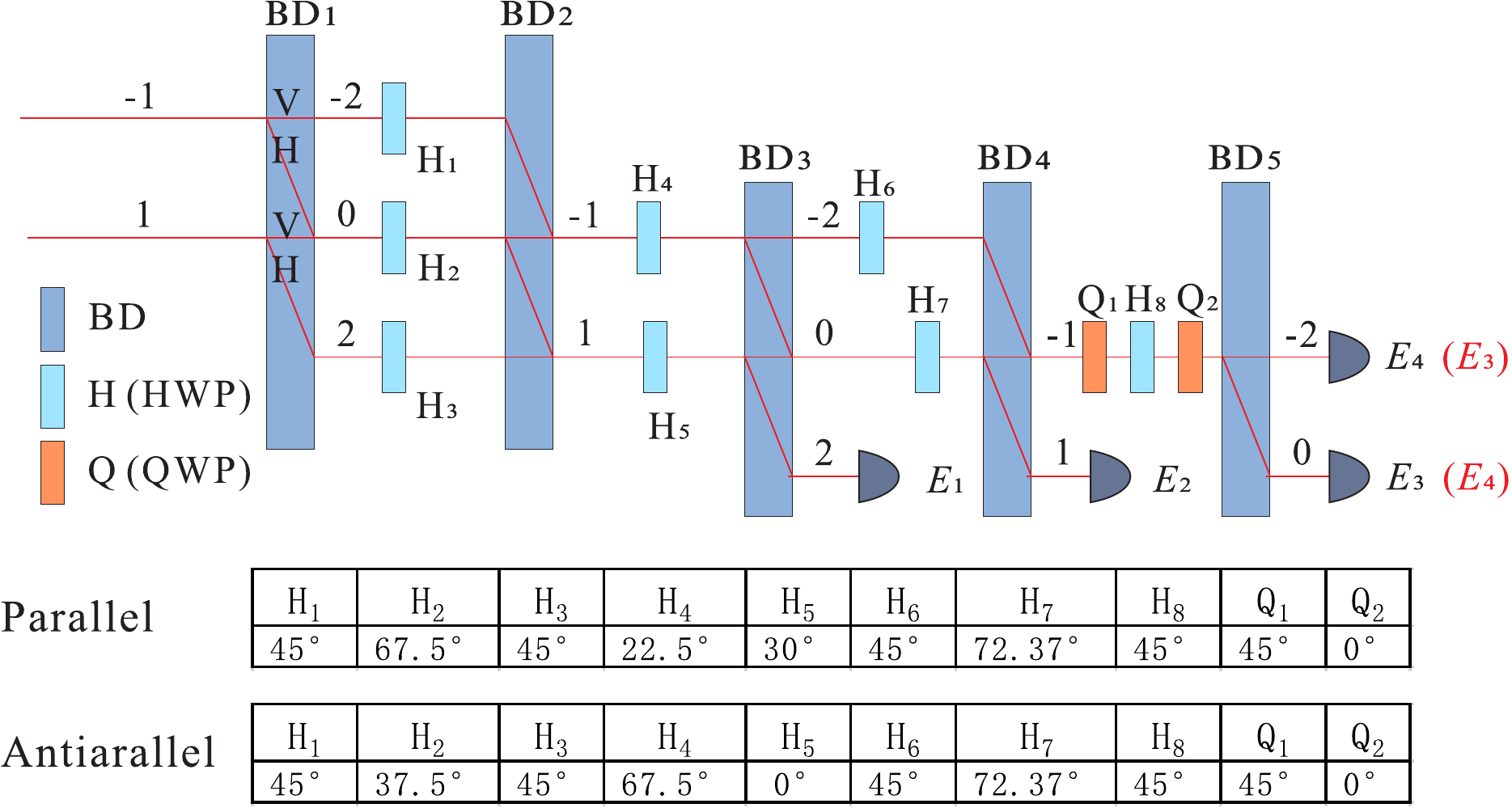}}
	\caption{\label{fig:para scheme} Realization of  the optimal entangling measurements  on parallel and antiparallel spins using photonic quantum walks. The translation operator is realized by beam displacers (BDs).  The nontrivial coin operators are realized by half wave plates (HWPs) and quarter wave plates (QWPs) with rotation angles specified in the tables embedded in the figure for parallel and antiparallel spins, respectively. The positions of the detectors $E_3$ and $E_4$ are switched for antiparallel spins  compared with parallel spins as marked in red.
	}
\end{figure}

The optimal entangling measurement on parallel spins presented in  \eref{eq:parallel spins} in the main text
can be realized via five-step quantum walks illustrated in \fref{fig:para scheme}. Here 
the nontrivial coin operators read
\begin{equation}\label{eq:coin operators parallel}
\begin{aligned}
&C(H_1)=C(H_3)=C(H_6)=\left(
\begin{array}{cc}
0 & 1 \\
1 & 0 \\
\end{array}
\right)\!, \quad
C(H_2)=\frac{1}{\sqrt{2}}\left(
\begin{array}{rr}
-1 & 1 \\
1 & 1 \\
\end{array}
\right)\!, \quad C(H_5)=\frac{1}{2}\left(
\begin{array}{cc}
1 & \sqrt{3} \\
\sqrt{3} & -1 \\
\end{array}
\right)\!,\\
&C(H_4)=\frac{1}{\sqrt{2}}\left(
\begin{array}{rr}
1 & 1 \\
1 & -1 \\
\end{array}
\right)\!, \quad
C(H_7)=\frac{1}{\sqrt{3}}\left(
\begin{array}{cc}
- \sqrt{2} & 1 \\
1 &  \sqrt{2} \\
\end{array}
\right)\!, \quad
C(Q_2H_8Q_1)=\frac{1-\rmi}{2}\left(
\begin{array}{rr}
1 & \rmi \\
-1 & \rmi \\
\end{array}
\right)\!.
\end{aligned}
\end{equation}
These coin operators can be realized by half wave plates (HWPs), quarter wave plates (QWPs), or their combinations. The unitary operators associated with  a HWP and a QWP with rotation angles $H, Q$ are respectively given by
\begin{equation}\label{eq:HQ}
C(H)=\sin (2H)\sigma_x+\cos(2H)\sigma_z,\qquad C(Q)=\frac{1+\rmi}{2}\left\{I-\rmi[\sin (2Q)\sigma_x+\cos(2Q)\sigma_z]\right\}.
\end{equation}
The unitary operator associated with a combination of HWPs and QWPs are the product of respective unitary operators; for example, $C(Q_2H_8Q_1)=C(Q_2)C(H_8)C(Q_1)$. It is easy to check that the nontrivial coin operators in \eref{eq:coin operators parallel} can be realized by HWPs and QWPs with rotation angles specified in the first table embedded in \fref{fig:para scheme}.
To verify the efficacy of this scheme, we shall investigate the evolution of a general pure state under the actions of coin operators and translation operator.

%
%
%
%
%

%

Suppose the initial walker-coin state has the form
\begin{equation}\label{eq:general pure state}
\ket{\Phi_0}=a\ket{1,H}+b\ket{1,V}+c\ket{-1,H}+d\ket{-1,V},
\end{equation}
where $a, b, c,d$ are complex coefficients satisfying the normalization condition $|a|^2 +|b|^2+|c|^2+|d|^2=1$. 
In the first step ($t=1$), all coin operators are trivial; under the action of BD$_1$, 
 the state $\ket{\Phi_0}$ evolves into
\begin{equation}
\ket{\Phi_1}=a\ket{2,H}+b\ket{0,V}+c\ket{0,H}+d\ket{-2,V}\,.
\end{equation}
In the second step ($t=2$), the nontrivial coin operators are generated by HWPs H$_1$, H$_2$, and  H$_3$, under the action of these wave plates and BD$_2$, the state $\ket{\Phi_1}$  evolves into
\begin{equation}
\ket{\Phi_2}=a\ket{1,V}+c^\prime\ket{1,H}+b^\prime\ket{-1,V}+d\ket{-1,H},
\end{equation}
where
\begin{equation*}
b^\prime\equiv \frac{b+c}{\sqrt{2}},\qquad c^\prime\equiv \frac{b-c}{\sqrt{2}}\,.
\end{equation*}
By the same token, after steps $t=3,4,5$, the respective walker-coin states  read
\begin{align}
\ket{\Phi_3}&=e_1\ket{2,H} + \frac{1}{2}\left(-a+\sqrt{3}c^\prime\right)\ket{0,V}+\frac{\sqrt{2}}{2}\left(b^\prime+d\right)\ket{0,H} - \frac{\sqrt{2}}{2}\left(b^\prime-d\right)\ket{-2,V},\\
\ket{\Phi_4}&=e_1\ket{3,H} +e_2\ket{1,H} +\frac{\sqrt{6}}{6}\left(-a+\sqrt{3}b^\prime+c^\prime+d\right)\ket{-1,V}+ \frac{\sqrt{2}}{2}(-b^\prime+d)\ket{-1,H},\\
\ket{\Phi_5}&=e_1\ket{4,H} +e_2\ket{2,H}+e_3\ket{0,H} +e_4\ket{-2,V},\label{eq:Phi5Para}
\end{align}
where
\begin{align*}
e_1&=\frac{1}{2}\left(\sqrt{3}a+c^\prime\right), & 
e_2&=-\frac{\sqrt{3}}{6}\left(a+2b^\prime-\sqrt{3}c^\prime+2d\right),\\
e_3&=-\frac{\sqrt{3}}{6}\rme^{\frac{\pi}{4}\rmi}\Bigl(a+2\rme^{-\frac{2\pi}{3}\rmi}b^\prime-{\sqrt{3}}c^\prime+2\rme^{-\frac{4\pi}{3}\rmi}d\Bigr),& 
e_4&=-\frac{\sqrt{3}}{6}\rme^{\frac{\pi}{4}\rmi} \Bigl(a+2\rme^{\frac{2\pi}{3}\rmi}b^\prime-{\sqrt{3}}c^\prime+2\rme^{\frac{4\pi}{3}\rmi}d\Bigr).
\end{align*}


Now the desired entangling measurement presented in \eref{eq:parallel spins} can be realized by measuring the walk position in view of \eref{eq:Phi5Para} and the following relations,
\begin{equation}
\begin{aligned}
|\inner{4,H}{\Phi_5}|^2=&\bigl|e_1\bigr|^2=\bigl|\inner{\Psi_1^{\parallel}}{\Phi_0}\bigr|^2,&
|\inner{2,H}{\Phi_5}|^2&=\bigl|e_2\bigr|^2=\bigl|\inner{\Psi_2^{\parallel}}{\Phi_0}\bigr|^2,\\
|\inner{0,H}{\Phi_5}|^2=&\bigl|e_3\bigr|^2=\bigl|\inner{\Psi_3^{\parallel}}{\Phi_0}\bigr|^2,&
|\inner{-2,V}{\Phi_5}|^2&=\bigl|e_4\bigr|^2=\bigl|\inner{\Psi_4^{\parallel}}{\Phi_0}\bigr|^2. 
\end{aligned}
\end{equation}
Notably, the positions $4, 2, 0, -2$ correspond to the four outcomes $\ket{\Psi_1^{\parallel}}$,  $\ket{\Psi_2^{\parallel}}$, $\ket{\Psi_3^{\parallel}}$ to $\ket{\Psi_4^{\parallel}}$, respectively. In addition, the detector at position 4 (2) after step 5 can be replaced by a detector at position 2 (1) after step 3 (4) without modifying the effective measurement, as illustrated in \fref{fig:para scheme}. In a word, the detectors  $E_1$ to $E_4$ in \fref{fig:para scheme} correspond to the four outcomes  $\ket{\Psi_1^{\parallel}}$,  $\ket{\Psi_2^{\parallel}}$, $\ket{\Psi_3^{\parallel}}$ to $\ket{\Psi_4^{\parallel}}$, respectively.

\subsection{Optimal entangling measurement for antiparallel spins}

The optimal entangling measurement on antiparallel spins presented in \eref{eq:antiparallel spins} can also be realized using five-step quantum walks in a similar way to the case of  parallel spins as shown in \fref{fig:para scheme}. The main differences are the rotation angles of HWPs (see the second table embedded in the figure), which lead to the following coin operators:
\begin{equation}
\begin{aligned}
&C(H_1)=C(H_3)=C(H_6)=\left(
\begin{array}{cc}
0 & 1 \\
1 & 0 \\
\end{array}
\right)\!, \quad
C(H_2)=\left(
\begin{array}{rr}
\eta_0 & \eta_1 \\
\eta_1 & -\eta_0 \\
\end{array}
\right)\!, \quad C(H_5)=\left(
\begin{array}{cc}
1 & 0 \\
0 & -1 \\
\end{array}
\right)\!,\\
&C(H_4)=\frac{1}{\sqrt{2}}\left(
\begin{array}{rr}
-1 & 1 \\
1 & 1 \\
\end{array}
\right)\!, \quad
C(H_7)=\frac{1}{\sqrt{3}}\left(
\begin{array}{cc}
-\sqrt{2} &1 \\
1 &  \sqrt{2} \\
\end{array}
\right)\!, \quad
C(Q_2H_8Q_1)=\frac{1-\rmi}{{2}}\left(
\begin{array}{rr}
1 & \rmi \\
-1 & \rmi \\
\end{array}
\right),
\end{aligned}
\end{equation}
where $\eta_0=\frac{\sqrt{6}-\sqrt{2}}{4}$ and $\eta_1=\frac{\sqrt{6}+\sqrt{2}}{4}$.


Suppose the initial walker-coin state has the form of \eref{eq:general pure state}. After the first five steps, the respective walker-coin states read
\begin{align}
\ket{\Phi_1}&=a\ket{2,H}+b\ket{0,V}+c\ket{0,H}+d\ket{-2,V},\\
\ket{\Phi_2}&=a\ket{1,V}+\left(\eta_1b+\eta_0c\right)\ket{1,H}+\left(-\eta_0b+\eta_1c\right)\ket{-1,V}+d\ket{-1,H},\\
\ket{\Phi_3}&=e_1\ket{2,H} -a\ket{0,V}+\frac1{\sqrt{2}}\left(b^\prime-d\right)\ket{0,H} + \frac1{\sqrt{2}}\left(b^\prime+d\right)\ket{-2,V},\\
\ket{\Phi_4}&=e_1\ket{3,H}+e_2\ket{1,H}+ \frac{\sqrt{3}}{3}\left(-\sqrt{2}a+\frac1{\sqrt{2}}b^\prime-\frac1{\sqrt{2}}d\right)\ket{-1,V}+ \frac1{\sqrt{2}}(b^\prime+d)\ket{-1,H},\\
\ket{\Phi_5}&=e_1\ket{4,H} +e_2\ket{2,H}+e_3\ket{0,H} +e_4\ket{-2,V},\label{eq:Phi5AP}
\end{align}
where 
\begin{align*}
b^\prime&=-\eta_0b+\eta_1c,\quad
e_1=\eta_1b+\eta_0c,\quad
e_2=\frac{\sqrt{3}}{3}\left(-a-b^\prime+d\right),\\
e_3&=-\frac{\sqrt{3}}{3}\rme^{\frac{\pi}{4}\rmi}\left(a+\rme^{\frac{2\pi}{3}\rmi}b^\prime-\rme^{\frac{4\pi}{3}\rmi}d\right),\quad
e_4=- \frac{\sqrt{3}}{3}\rme^{\frac{\pi}{4}\rmi}\left(a+\rme^{-\frac{2\pi}{3}\rmi}b^\prime-\rme^{-\frac{4\pi}{3}\rmi}d\right).
\end{align*}

Now the desired entangling measurement presented in \eref{eq:antiparallel spins} can be realized by measuring the walk position in view of \eref{eq:Phi5AP} and the following relations,
\begin{equation}
\begin{aligned}
|\inner{4,H}{\Phi_5}|^2=&\left|e_1\right|^2=\left|\inner{\Psi_1^{\perp}}{\Phi_0}\right|^2,&
|\inner{2,H}{\Phi_5}|^2&=\left|e_2\right|^2=\left|\inner{\Psi_2^{\perp}}{\Phi_0}\right|^2,\\
|\inner{0,H}{\Phi_5}|^2=&\left|e_3\right|^2=\left|\inner{\Psi_4^{\perp}}{\Phi_0}\right|^2,&
|\inner{-2,V}{\Phi_5}|^2&=\left|e_4\right|^2=\left|\inner{\Psi_3^{\perp}}{\Phi_0}\right|^2.
\end{aligned}
\end{equation}
Notably, the positions $4, 2, 0, -2$ correspond to the four outcomes $\ket{\Psi_1^{\perp}}$,  $\ket{\Psi_2^{\perp}}$, $\ket{\Psi_4^{\perp}}$ to $\ket{\Psi_3^{\perp}}$, respectively.   In addition, the detector at position 4 (2) after step 5 can be replaced by a detector at position 2 (1) after step 3 (4) without modifying the effective measurement, as illustrated in \fref{fig:para scheme}. In a word, the detectors  $E_1$ to $E_4$ in \fref{fig:para scheme} correspond to the four outcomes  $\ket{\Psi_1^{\perp}}$,  $\ket{\Psi_2^{\perp}}$, $\ket{\Psi_3^{\perp}}$ to $\ket{\Psi_4^{\perp}}$, respectively. Note that the positions of the detectors $E_3$ and $E_4$ (marked in red) are switched compared with the case of  parallel spins.


\subsection{Local projective  measurement $\sigma_x\sigma_y$}

\begin{figure}[t]
	\center{\includegraphics[scale=0.6]{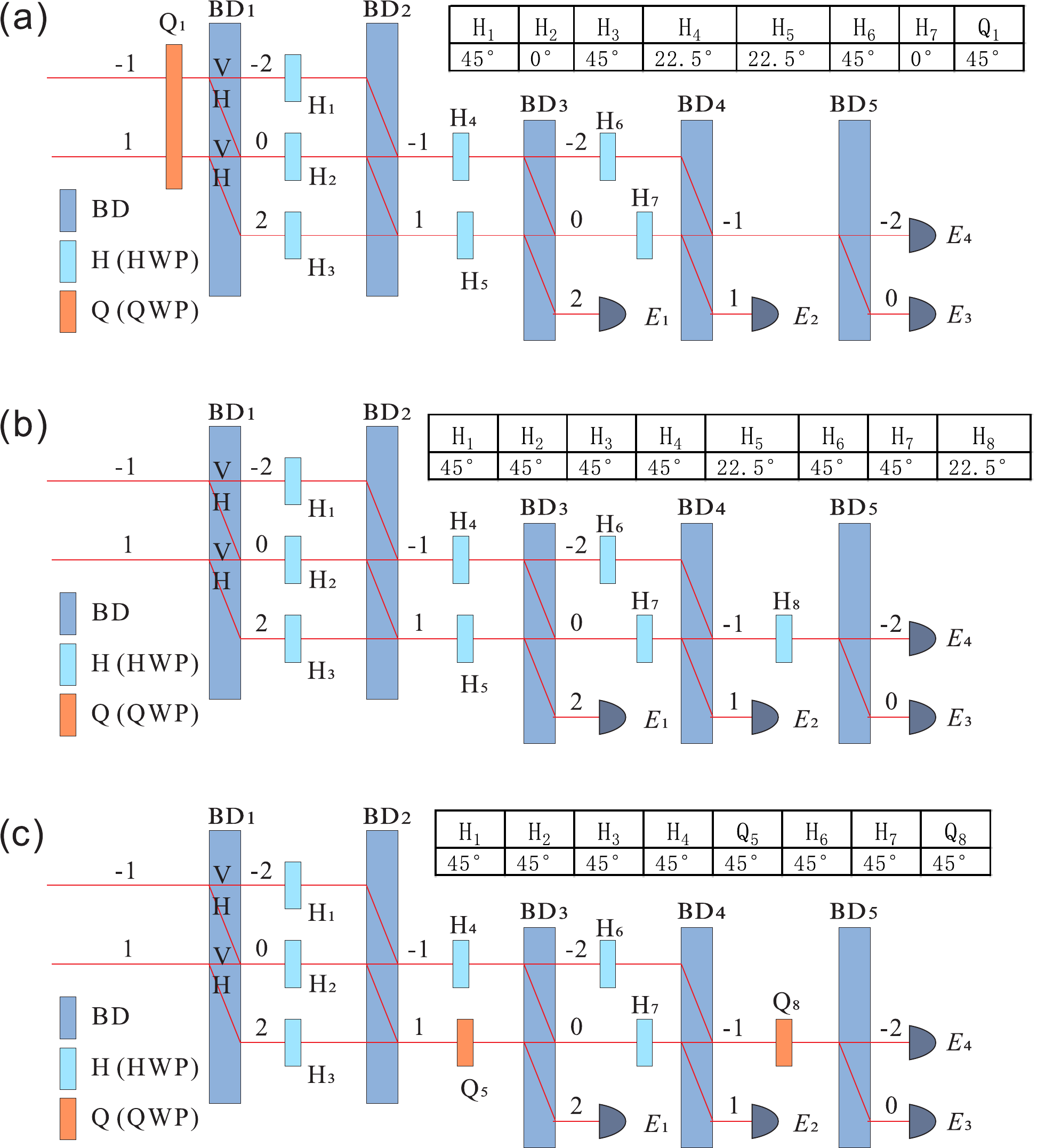}}
	\caption{\label{fig:XYscheme}  Realization of  three local projective  measurements on parallel spins via photonic quantum walks: (a) $\sigma_x\sigma_y$ (b) $\sigma_z\sigma_x$ (c) $\sigma_z\sigma_y$. The nontrivial coin operators are realized using HWPs and QWPs with rotation angles specified in the three  tables embedded.  
	}
\end{figure}

The local projective measurement  $\sigma_x\sigma_y$ has four outcomes corresponding to the four projectors $E_i=\outer{\Psi_i}{\Psi_i}$, with  $\ket{\Psi_1}=\ket{+_x+_y}$, $\ket{\Psi_2}=\ket{-_x-_y}$, $\ket{\Psi_3}=\ket{+_x-_y}$, $\ket{\Psi_4}=\ket{-_x+_y}$. Here $\ket{\pm_{x(y)}}$ are eigenvectors of the Pauli operator $\sigma_{x(y)}$ corresponding to eigenvalues $\pm1$. This measurement can also be realized using five-step photonic quantum walks as shown in plot (a) of  \fref{fig:XYscheme}. The  nontrivial coin operators are
\begin{equation}
\begin{aligned}
C(H_1)&=C(H_3)=C(H_6)=\left(
\begin{array}{cc}
0 & 1 \\
1 & 0 \\
\end{array}
\right)\!, &
C(H_2)&=C(H_7)\left(
\begin{array}{rr}
1 & 0 \\
0 & -1 \\
\end{array}
\right)\!,\\
C(H_4)&=C(H_5)=\frac{1}{\sqrt{2}}\left(
\begin{array}{rr}
1 & 1 \\
1 & -1 \\
\end{array}
\right)\!, &
C(Q_1)&=\frac{1+\rmi}{2}\left(
\begin{array}{rr}
1 & -\rmi \\
-\rmi & 1 \\
\end{array}
\right)\!.
\end{aligned}
\end{equation}
These coin operators can be realized by wave plates with rotation angles specified in the first table embedded in \fref{fig:XYscheme}.

In the first five steps, the initial walker-coin state in  \eref{eq:general pure state} evolves into
\begin{align}
\ket{\Phi_1}&=a^\prime\ket{2,H}+b^\prime\ket{0,V}+c^\prime\ket{0,H}+d^\prime\ket{-2,V},\\
\ket{\Phi_2}&=a^\prime\ket{1,V}+c^\prime\ket{1,H}-b^\prime\ket{-1,V}+d^\prime\ket{-1,H},\\
\ket{\Phi_3}&=e_1\ket{2,H} -\frac1{\sqrt{2}}\left(a^\prime-c^\prime\right)\ket{0,V}-\frac1{\sqrt{2}}\left(b^\prime-d^\prime\right)\ket{0,H}+ \frac1{\sqrt{2}}\left(b^\prime+d^\prime\right)\ket{-2,V},\\
\ket{\Phi_4}&=e_1\ket{3,H}+e_2\ket{1,H}+ \frac1{\sqrt{2}}\left(a^\prime-c^\prime\right)\ket{-1,V}+\frac1{\sqrt{2}}\left(b^\prime+d^\prime\right)\ket{-1,H},\\
\ket{\Phi_5}&=e_1\ket{4,H} +e_2\ket{2,H}+e_3\ket{0,H} +e_4\ket{-2,V},\label{eq:Phixy}
\end{align}
where
\begin{align*}
a^\prime&= \frac{\rme^{\frac{\pi}{4}\rmi}}{\sqrt{2}}(a-\rmi b), & b^\prime&=  \frac{\rme^{\frac{\pi}{4}\rmi}}{\sqrt{2}}(-\rmi a+b),&  c^\prime&=  \frac{\rme^{\frac{\pi}{4}\rmi}}{\sqrt{2}}(c-\rmi d),& d^\prime&=  \frac{\rme^{\frac{\pi}{4}\rmi}}{\sqrt{2}}(-\rmi c+d),\\
e_1&=\frac1{\sqrt{2}}\left(a^\prime+c^\prime\right),& e_2&=-\frac1{\sqrt{2}}\left(b^\prime-d^\prime\right),&
e_3&=\frac1{\sqrt{2}}\left(b^\prime+d^\prime\right),&
e_4&=\frac1{\sqrt{2}}\left(a^\prime-c^\prime\right).
\end{align*}

Now the local projective measurement $\sigma_x\sigma_y$  can be realized by measuring the walk position in view of \eref{eq:Phixy} and the following relations,
\begin{align}
|\inner{4,H}{\Phi_5}|^2=&\left|e_1\right|^2=\left|\inner{+_x,+_y}{\Phi_0}\right|^2,&
|\inner{2,H}{\Phi_5}|^2&=\left|e_2\right|^2=\left|\inner{-_x,-_y}{\Phi_0}\right|^2,\\
|\inner{0,H}{\Phi_5}|^2=&\left|e_3\right|^2=\left|\inner{+_x,-_y}{\Phi_0}\right|^2,&
|\inner{-2,V}{\Phi_5}|^2&=\left|e_4\right|^2=\left|\inner{-_x,+_y}{\Phi_0}\right|^2.
\end{align}
The detectors  $E_1$ to $E_4$ in plot (a)  of \fref{fig:XYscheme} correspond to the four outcomes $\ket{\Psi_1}$ to $\ket{\Psi_4}$, respectively.

\subsection{Local projective  measurement $\sigma_z\sigma_x$}

The local projective measurement  $\sigma_z\sigma_x$ has four outcomes corresponding to the four projectors $E_i=\outer{\Psi_i}{\Psi_i}$, with  $\ket{\Psi_1}=\ket{+_z+_x}$, $\ket{\Psi_2}=\ket{+_z-_x}$, $\ket{\Psi_3}=\ket{-_z+_x},\ket{\Psi_4}=\ket{-_z-_x}$. Here $\ket{\pm_{z(x)}}$ are eigenvectors of the Pauli operator $\sigma_{z(x)}$ corresponding to eigenvalues $\pm1$. This measurement can also be realized using five-step photonic quantum walks as shown in plot (b) of  \fref{fig:XYscheme}. The  nontrivial coin operators read
\begin{equation}
\begin{aligned}
&C(H_1)=C(H_2)=C(H_3)=C(H_4)=C(H_6)=C(H_7)=\left(
\begin{array}{cc}
0 & 1 \\
1 & 0 \\
\end{array}
\right)\!,\\
&C(H_5)=C(H_8)=\frac{1}{\sqrt{2}}\left(
\begin{array}{rr}
1 & 1 \\
1 & -1 \\
\end{array}
\right)\!.
\end{aligned}
\end{equation}

In the first five steps, the initial walker-coin state in  \eref{eq:general pure state} evolves into
\begin{align}
\ket{\Phi_1}&=a\ket{2,H}+b\ket{0,V}+c\ket{0,H}+d\ket{-2,V},\\
\ket{\Phi_2}&=a\ket{1,V}+b\ket{1,H}+c\ket{-1,V}+d\ket{-1,H},\\
\ket{\Phi_3}&=e_1\ket{2,H}-\frac1{\sqrt{2}}\left(a-b\right)\ket{0,V}+c\ket{0,H}+ d\ket{-2,V},\\
\ket{\Phi_4}&=e_1\ket{3,H}+e_2\ket{1,H}+ c\ket{-1,V}+d\ket{-1,H},\\
\ket{\Phi_5}&=e_1\ket{4,H} +e_2\ket{2,H}+e_3\ket{0,H} +e_4\ket{-2,V},\label{eq:Phi5zx}
\end{align}
where
\begin{equation}
e_1=\frac1{\sqrt{2}}\left(a+b\right),\quad
e_2=-\frac1{\sqrt{2}}\left(a-b\right),\quad
e_3=\frac1{\sqrt{2}}\left(c+d\right),\quad
e_4=-\frac1{\sqrt{2}}\left(c-d\right).
\end{equation}

Now the local projective measurement $\sigma_z\sigma_x$  can be realized by measuring the walk position in view of \eref{eq:Phi5zx} and the following relations,
\begin{align}
|\inner{4,H}{\Phi_5}|^2=&\left|e_1\right|^2=\left|\inner{+_z,+_x}{\Phi_0}\right|^2,&
|\inner{2,H}{\Phi_5}|^2&=\left|e_2\right|^2=\left|\inner{+_z,-_x}{\Phi_0}\right|^2,\\
|\inner{0,H}{\Phi_5}|^2=&\left|e_3\right|^2=\left|\inner{-_z,+_x}{\Phi_0}\right|^2,&
|\inner{-2,V}{\Phi_5}|^2&=\left|e_4\right|^2=\left|\inner{-_z,-_x}{\Phi_0}\right|^2.
\end{align}
The detectors  $E_1$ to $E_4$ in plot (b)  of \fref{fig:XYscheme} correspond to the four outcomes $\ket{\Psi_1}$ to  $\ket{\Psi_4}$, respectively.

\subsection{Local projective  measurement  $\sigma_z\sigma_y$}

The local projective measurement  $\sigma_z\sigma_y$ has four outcomes corresponding to the four projectors $E_i=\outer{\Psi_i}{\Psi_i}$, with  $\ket{\Psi_1}=\ket{+_z-_y}$, $\ket{\Psi_2}=\ket{+_z+_y}$, $\ket{\Psi_3}=\ket{-_z-_y},\ket{\Psi_4}=\ket{-_z+_x}$. Here $\ket{\pm_{z(y)}}$ are eigenvectors of the Pauli operator $\sigma_{z(y)}$ corresponding to eigenvalues $\pm1$. This measurement can also be realized using five-step photonic quantum walks as shown in plot (c) of  \fref{fig:XYscheme}. The  nontrivial coin operators read
\begin{equation}
\begin{aligned}
&C(H_1)=C(H_2)=C(H_3)=C(H_4)=C(H_6)=C(H_7)=\left(
\begin{array}{cc}
0 & 1 \\
1 & 0 \\
\end{array}
\right)\!,\\
&C(Q_5)=C(Q_8)=\frac{1+\rmi}{2}\left(
\begin{array}{rr}
1 & -\rmi \\
-\rmi & 1 \\
\end{array}
\right)\!,
\end{aligned}
\end{equation}

In the first five steps, the initial walker-coin state in  \eref{eq:general pure state} evolves into
\begin{align}
\ket{\Phi_1}&=a\ket{2,H}+b\ket{0,V}+c\ket{0,H}+d\ket{-2,V},\\
\ket{\Phi_2}&=a\ket{1,V}+b\ket{1,H}+c\ket{-1,V}+d\ket{-1,H},\\
\ket{\Phi_3}&=e_1\ket{2,H}+\frac{\rme^{\frac{\pi}{4}\rmi}}{\sqrt{2}}\left(a-\rmi b\right)\ket{0,V}+c\ket{0,H}+ d\ket{-2,V},\\
\ket{\Phi_4}&=e_1\ket{3,H}+e_2\ket{1,H}+ c\ket{-1,V}+d\ket{-1,H},\\
\ket{\Phi_5}&=e_1\ket{4,H} +e_2\ket{2,H}+e_3\ket{0,H} +e_4\ket{-2,V},\label{eq:Phi5zy}
\end{align}
where
\begin{equation}
e_1=\frac{\rme^{-\frac{\pi}{4}\rmi}}{\sqrt{2}}\left(a+\rmi b\right),\quad
e_2=\frac{\rme^{\frac{\pi}{4}\rmi}}{\sqrt{2}}\left(a-\rmi b\right),\quad
e_3=\frac{\rme^{-\frac{\pi}{4}\rmi}}{\sqrt{2}}\left(c+\rmi d\right),\quad
e_4=\frac{\rme^{\frac{\pi}{4}\rmi}}{\sqrt{2}}\left(c-\rmi d\right).
\end{equation}

Now the local projective measurement $\sigma_z\sigma_y$  can be realized by measuring the walk position in view of \eref{eq:Phi5zy} and the following relations,
\begin{align}
|\inner{4,H}{\Phi_5}|^2=&\left|e_1\right|^2=\left|\inner{+_z,-_y}{\Phi_0}\right|^2,&
|\inner{2,H}{\Phi_5}|^2&=\left|e_2\right|^2=\left|\inner{+_z,+_y}{\Phi_0}\right|^2,\\
|\inner{0,H}{\Phi_5}|^2=&\left|e_3\right|^2=\left|\inner{-_z,-_y}{\Phi_0}\right|^2,&
|\inner{-2,V}{\Phi_5}|^2&=\left|e_4\right|^2=\left|\inner{-_z,+_y}{\Phi_0}\right|^2.
\end{align}
The detectors  $E_1$ to $E_4$ in plot (c)  of \fref{fig:XYscheme} correspond to the four outcomes $\ket{\Psi_1}$ to $\ket{\Psi_4}$, respectively.

\bigskip
\bigskip
\section{Experimental measurement tomography}
\begin{figure*}[b]
	\center{\includegraphics[scale=0.55]{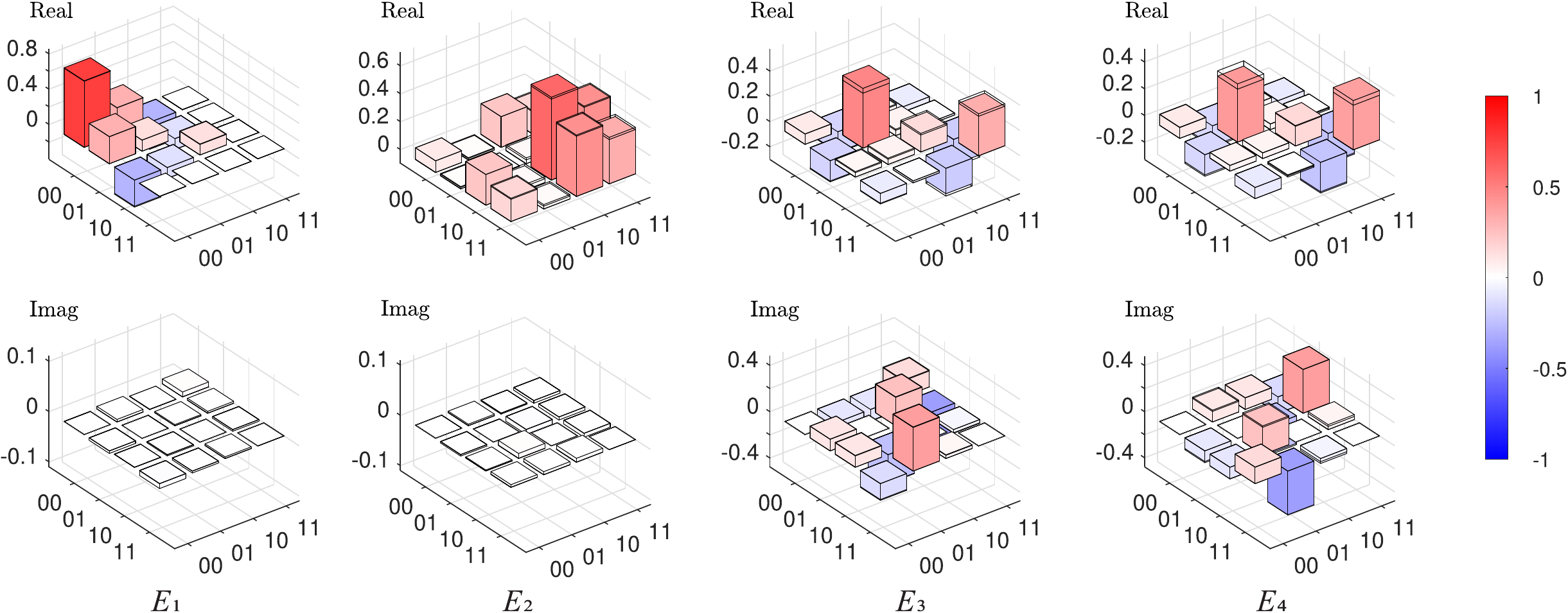}}
	\caption{\label{fig:tomo parallel} Results on measurement tomography of the optimal  entangling measurement  for parallel spins in \eref{eq:parallel spins} realized using photonic quantum walks. The matrix elements of the real (Real) and imaginary (Imag) parts of the four projectors $E_1$ to $E_4$ are
	plotted with solid colors.  For comparison, the counterparts of the ideal projectors are plotted as wire frames.}
\end{figure*}

\begin{figure*}[t]
	\center{\includegraphics[scale=0.55]{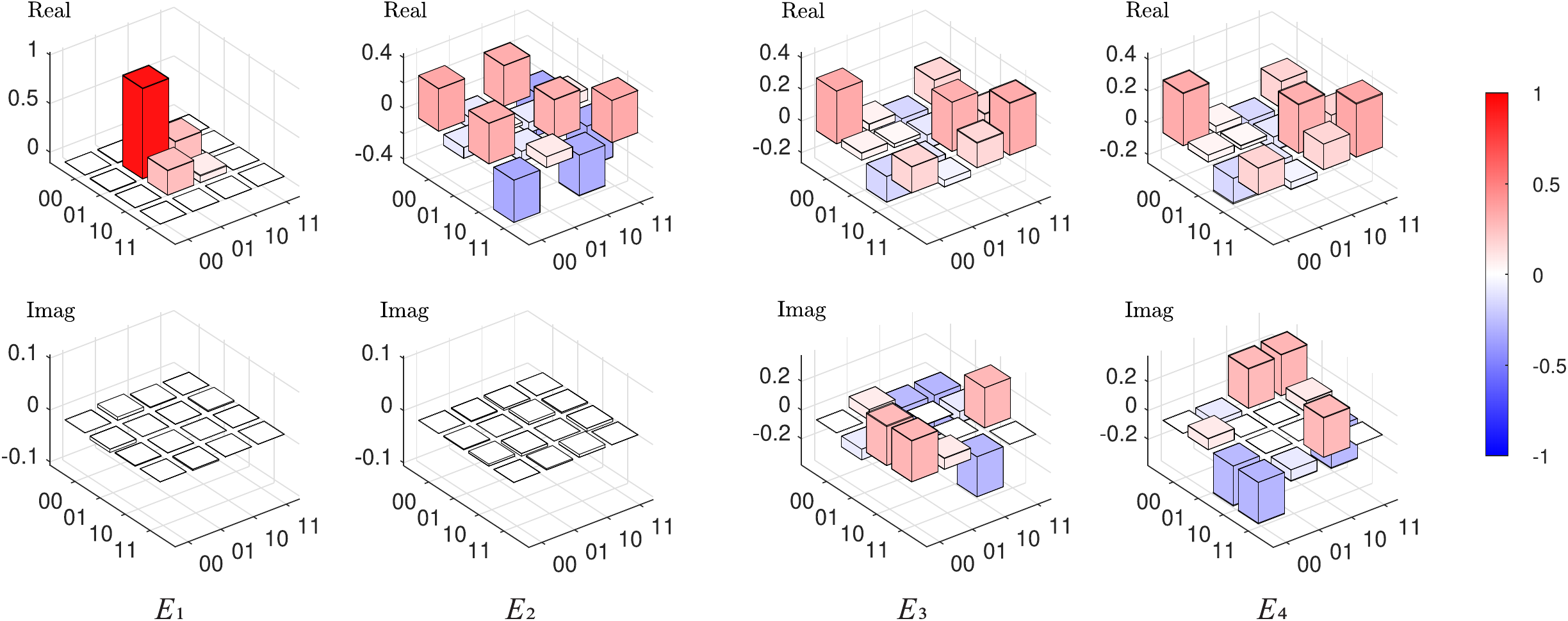}}
	\caption{\label{fig:tomo anti}
Results on measurement tomography of the optimal  entangling measurement  for antiparallel spins in \eref{eq:antiparallel spins} realized using photonic quantum walks. The matrix elements of the real (Real) and imaginary (Imag) parts of the four projectors $E_1$ to $E_4$ are
plotted with solid colors.  For comparison, the counterparts of the ideal projectors are plotted as wire frames.		
}
\end{figure*}

\begin{figure*}
\center{\includegraphics[scale=0.55]{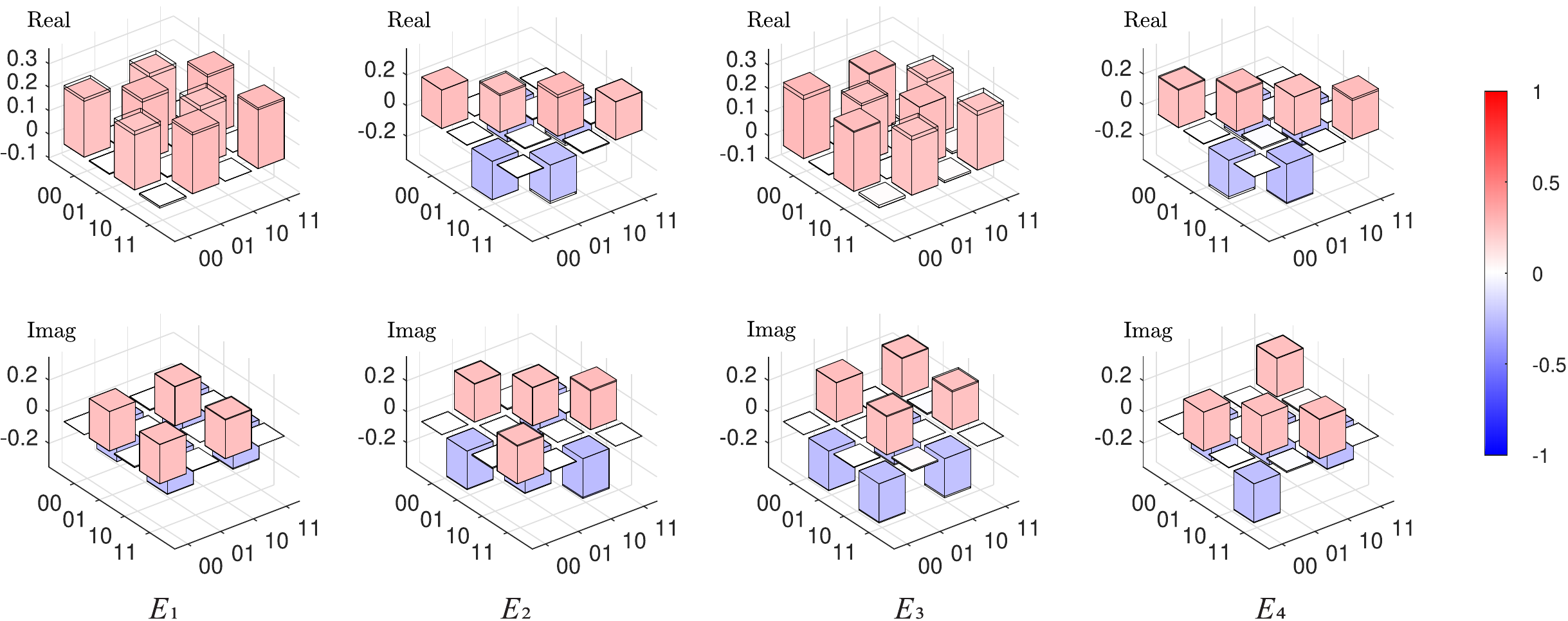}}
\caption{\label{fig:tomo XY}
Results on measurement tomography of the local projective measurement $\sigma_x\sigma_y$ on parallel spins realized using photonic quantum walks. The matrix elements of the real (Real) and imaginary (Imag) parts of the four projectors $E_1$ to $E_4$ are
plotted with solid colors.  For comparison, the counterparts of the ideal projectors are plotted as wire frames.	
}
\end{figure*}

\begin{figure*}
	\center{\includegraphics[scale=0.55]{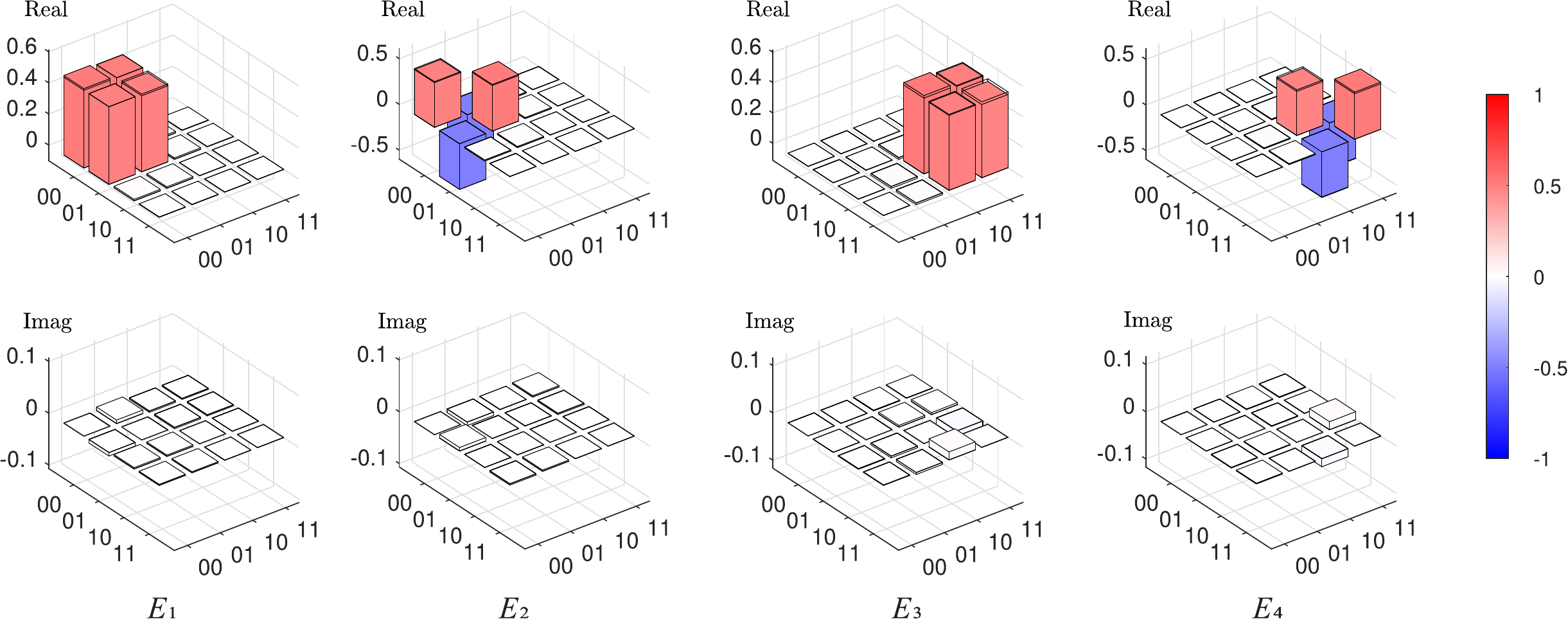}}
	\caption{\label{fig:tomo ZX}
Results on measurement tomography of the local projective measurement $\sigma_z\sigma_x$ on parallel spins realized using photonic quantum walks. The matrix elements of the real (Real) and imaginary (Imag) parts of the four projectors $E_1$ to $E_4$ are
plotted with solid colors.  For comparison, the counterparts of the ideal projectors are plotted as wire frames.	
	}
\end{figure*}

\begin{figure*}
	\center{\includegraphics[scale=0.55]{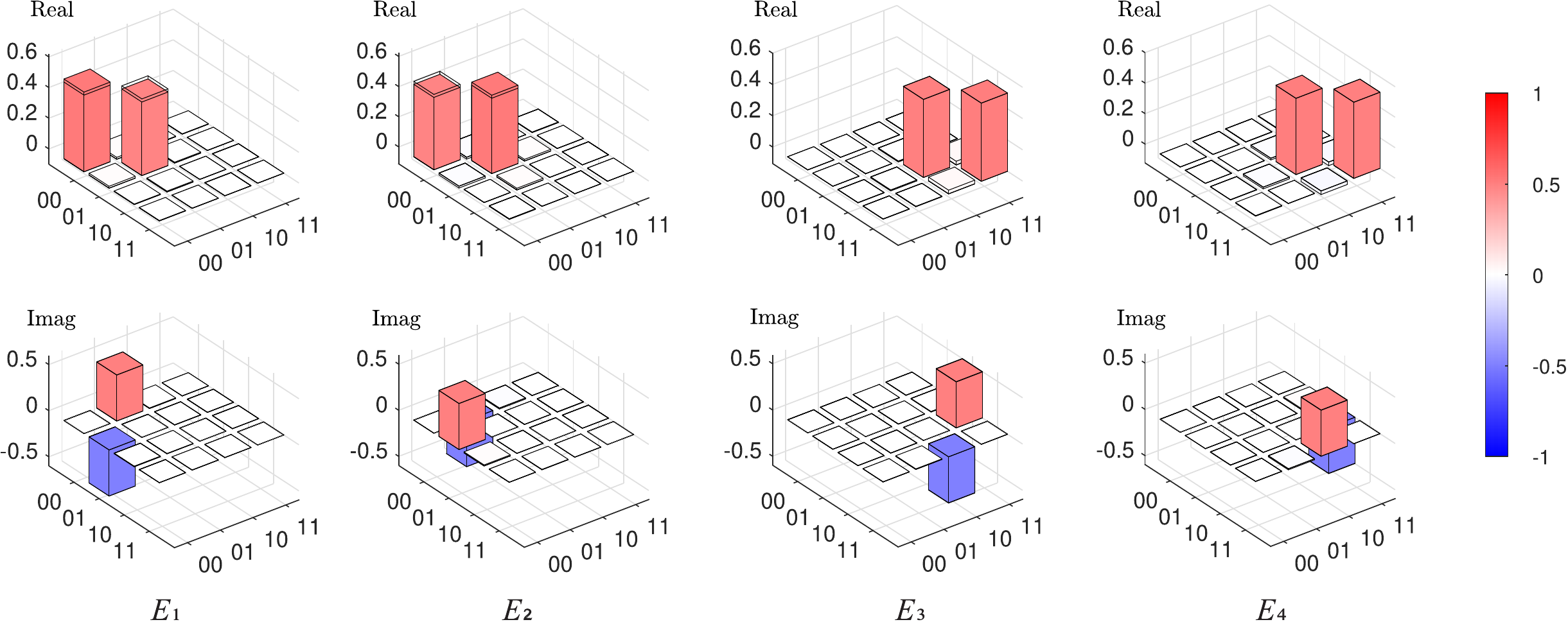}}
	\caption{\label{fig:tomo ZY}
Results on measurement tomography of the local projective measurement $\sigma_z\sigma_y$ on parallel spins realized using photonic quantum walks. The matrix elements of the real (Real) and imaginary (Imag) parts of the four projectors $E_1$ to $E_4$ are
plotted with solid colors.  For comparison, the counterparts of the ideal projectors are plotted as wire frames.	
	}
\end{figure*}

In this section, we provide more details on the measurement tomography of the five projective measurements realized using photonic quantum walks. Two of them are the optimal entangling measurements for parallel and antiparallel spins, respectively, while the other three are optimal local projective measurements. To perform measurement tomography, 36 states,  the tensor products of the six eigenstates of three Pauli operators, were prepared
and sent to the measurement module. To reduce statistical fluctuation, each state was prepared and measured
100000 times. Then the four projectors were estimated from the  measurement statistics using the maximum likelihood (ML) method developed in \rcite{Fiur01maximum}. The  projectors reconstructed are shown in \fsref{fig:tomo parallel} to \ref{fig:tomo ZY} for two entangling measurements and three  local projective  measurements, respectively. The fidelity of each projector and overall fidelity of each projective measurement are presented in \tref{measurement fideltiy}.   These results show that all the five projective measurements were realized with very high qualities.

%

\begin{table*}[b]
	\caption{\label{measurement fideltiy}Fidelities of two entangling measurements and three local projective measurements realized using photonic quantum walks (cf.~\tref{tab:dictionary}). Here  measurement tomography is employed to estimate the fidelity of each projector and the overall fidelity of each measurement.  Uncertainty of the fidelity in the parentheses denotes the standard deviation of 100 simulations from Poisson statistics.
	}
	\begin{tabular*}{1\textwidth}{@{\extracolsep{\fill}} c| c c c c c}
		\hline\hline
		Measurement schemes & $E_1$& $E_2$&$E_3$ & $E_4$& overall\\
		\hline
		parallel&0.9971(4)&0.9906(4)&0.9929(3)&0.9857(4)&0.9916(2)\\
		antiparallel&0.9974(3)&0.9916(4)&0.9911(4)&0.9907(4)&0.9927(2)\\
		\hline
		$\sigma_x\sigma_y$&0.9884(2)&0.9919(2)&0.9820(2)&0.9986(1)&0.9902(1)\\
		$\sigma_z\sigma_x$&0.9989(1)&0.9987(1)&0.9938(2)&0.9946(1)&0.9965(1)\\
		$\sigma_z\sigma_y$&0.9984(1)&0.9986(1)&0.9960(1)&0.9961(1)&0.9973(1)\\
		\hline\hline
	\end{tabular*}
\end{table*}

\end{document}